\documentclass[twocolumn,prl]{revtex4}
\usepackage{graphicx}
\usepackage{dcolumn}
\usepackage{citesort}
\usepackage{amsmath}
\usepackage{bm}
\begin{document}
\title{Analogy of the slow dynamics
between the supercooled liquid and the supercooled plastic crystal states of difluorotetrachloroethane}
\author{F.~Affouard and M.~Descamps}
\affiliation{
Laboratoire de Dynamique et Structure des Mat\'eriaux Mol\'eculaires,\\
CNRS UMR 8024,
Universit\'e Lille I,\\
59655 Villeneuve d'Ascq Cedex France}
\date{\today}
\begin{abstract}
Slow dynamics of difluorotetrachloroethane in both supercooled plastic crystal
and supercooled liquid states
have been investigated from Molecular Dynamics simulations. 
The temperature and wave-vector dependence of collective dynamics 
in both states are probed using coherent 
dynamical scattering functions $S(Q,t)$.
Our results confirm the strong analogy between molecular liquids and plastic crystals
for which $\alpha$-relaxation times and 
non-ergodicity parameters are controlled by the non trivial static correlations $S(Q)$
as predicted by the Mode Coupling Theory. The use of  
infinitely thin needles distributed on a lattice as model of plastic crystals 
is discussed.
\end{abstract}

\maketitle

It is well known that very different materials such as:
silica, low-molecular-weight liquids (carbohydrates, alcohols),
polymers or even proteins are able to
exhibit a very intriguing feature called \emph{glass transition}~\cite{Debenedetti_nature01}. 
This latter is characterized by an extraordinary
decrease of several orders of magnitude
of the mobility in a narrow temperature range.
Translational, orientational or even internal degrees of freedom (TDOF, ODOF or IDOF) can be at the origin of the mobility 
of a specific substance and approaching the glass transition transition temperature $T_{g}$, all of them are inextricably coupled.
Most of the investigations performed on glassy  materials considered molecular liquids 
and focused on TDOF so far. 
The precise role of the ODOF remains particularly unclear as demonstrated by the so-called low-temperature
translation-rotation paradox~\cite{Blackburn_jpc96}.
Plastic crystals \emph{i.e}  molecular crystals composed of orientationally disordered molecules
offer an interesting
solution in order to mainly investigate ODOF during the freezing process.
Indeed, some of the plastic crystals called \emph{glassy crystals}~\cite{Suga_jncs74}
such as cyanoadamantane~\cite{Affouard_jncs02}, ethanol~\cite{Benkhof_jpcm98}  or
orthocarborane~\cite{Winterlich_prl03}
can be considered are true rotational analogs of canonical liquid glassformers since they show 
a step of the specific heat at $T_{g}$
or a non-Arrhenius behaviour
of the rotational relaxation times.
Only a few substances show the extraordinary property to exhibit 
a glass transition 
from both the plastic and the liquid phase:
ethanol~\cite{Benkhof_jpcm98,Miller_prb98}, cyclohexene~\cite{Haida_bcsj77}, PMS~\cite{Fujimori_jncs96} or
2-bromothiophene~\cite{Fujimori_jpcs93}. An investigation of such materials is particularly interesting to understand 
the fundamental microscopic mechanisms of the glass transition and the precise interplay between the different degrees of freedom.

In the high temperatures pico-nanosecond regime (ps-ns), a fundamental question
concerns the onset of the precursor cooperative mechanisms which lead to the glass transition at $T_{g}$.
In the last two decades, much attention has been devoted to the mode coupling
theory (MCT)~\cite{Goetze}. So far, it is the only theory which provides a
microscopic description of supercooled atomic liquids. The intrinsic basis of MCT states that
the behaviour of any time-dependent correlators describing the dynamics of the system is only controlled
by its static density correlator \emph{i.e} $S(Q)$ and predicts the existence of a
critical temperature $T_{c}$.
Scaling properties predicted by MCT approaching
$T_{c}$ have been successfully validated in numerous experiments and molecular dynamics (MD) simulations~\cite{Schilling_03}
of systems whose dynamics are controlled by TDOF.
Recently, some extensions of MCT called molecular mode coupling theory (MMCT)~\cite{Schilling_03}
have been proposed to take ODOF into account:
one diatomic probe molecule in an atomic liquid, liquids made of linear molecules or
water to cite only a few. The authors have particularly shown that some of the basic
predictions of MCT still hold owing TDOF/ODOF coupling. As revealed by
a recent MD investigations~\cite{Chong_epl03} performed on orthoterphenyl (OTP),
coupling of the rotational dynamics to the center-of-mass motion can be very complex.
No microscopic theory has been developed for plastic crystals so far. However,
in~\cite{Affouard_jncs02,Affouard_prl01},
we have particularly shown from NMR and Raman experiments, and
MD computer simulations that
some predictions of
the idealized version of the MCT (critical temperature $T_{c}$ and time scaling laws)
were able to describe
relatively well rotational dynamics of different plastic crystals.
This intriguing previous result call for new investigations
to clarify the similarity between slow dynamics behaviour of plastic crystals and  molecular liquids.

In this Letter, we present results
of a Molecular Dynamics (MD) 
comparative numerical study of the supercooled plastic and the
supercooled liquid phases of difluorotetrachloroethane (DFTCE).
It should be noted that the
supercooled state of liquid DFTCE is obtained
owing the hyperquenching rate of the MD simulation.
This compound is composed of simple molecules $\mathrm{CFCl_{2}-CFCl_{2}}$ close to
dumbbells extensively used in MD calculations as prototype
of molecular glass-formers liquids~\cite{Kaemmerer_pre97}.
DFTCE plastic crystal has been experimentally widely studied
and exhibits a glass transition of the overall rotation of the
molecules at $T_{g} = 86 $ K~\cite{Kruger_jpcm94}.
Changes concerning the nature of dynamics in this system
has been reported from NMR experiments, Brillouin and
dielectric spectroscopy where it was suggested
that the freezing process could be described on the basis of MCT~\cite{Kruger_jpcm94}.

\begin{figure}[h]
\includegraphics[width=7cm,clip=true]{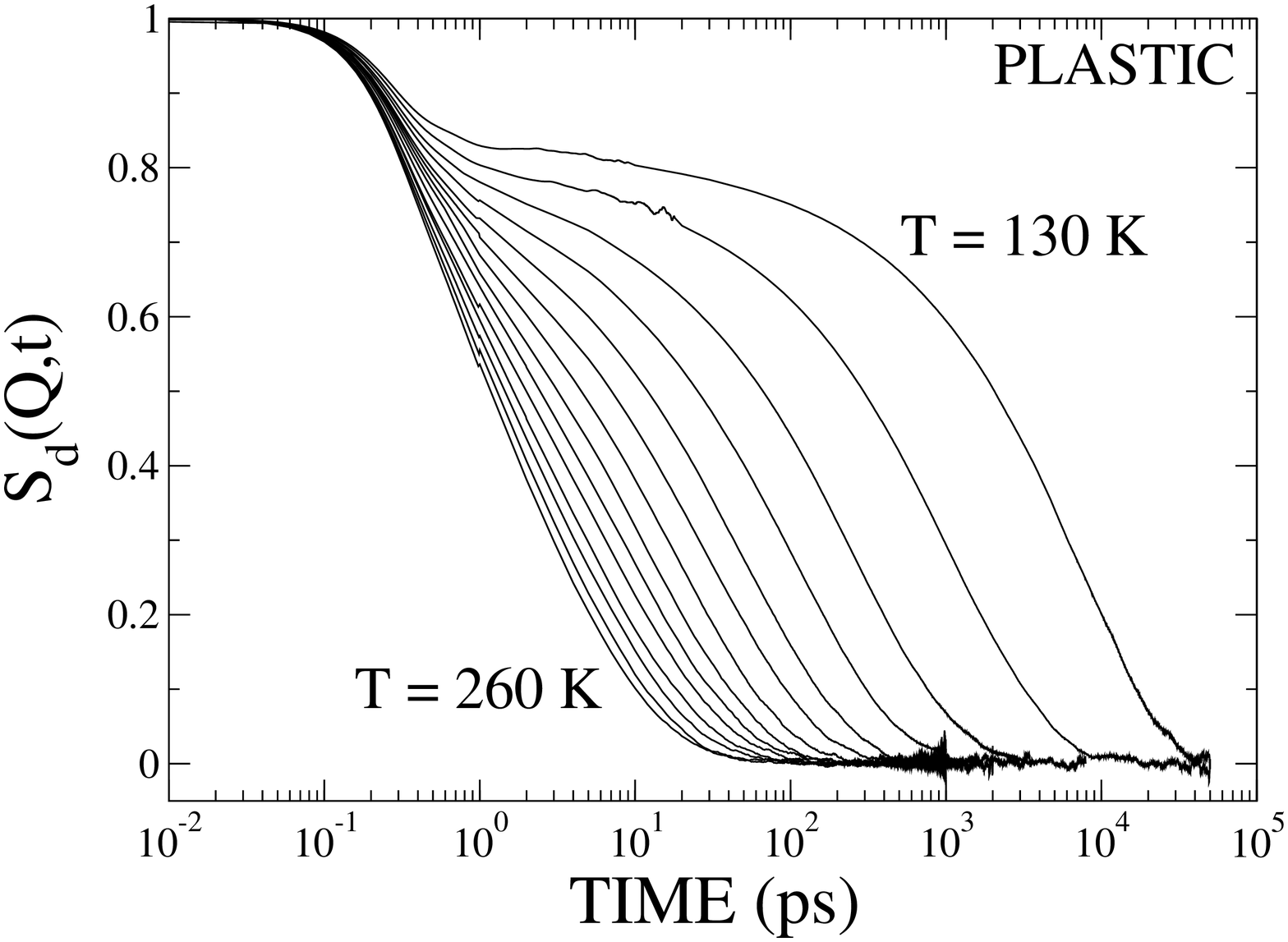}
\includegraphics[width=7cm,clip=true]{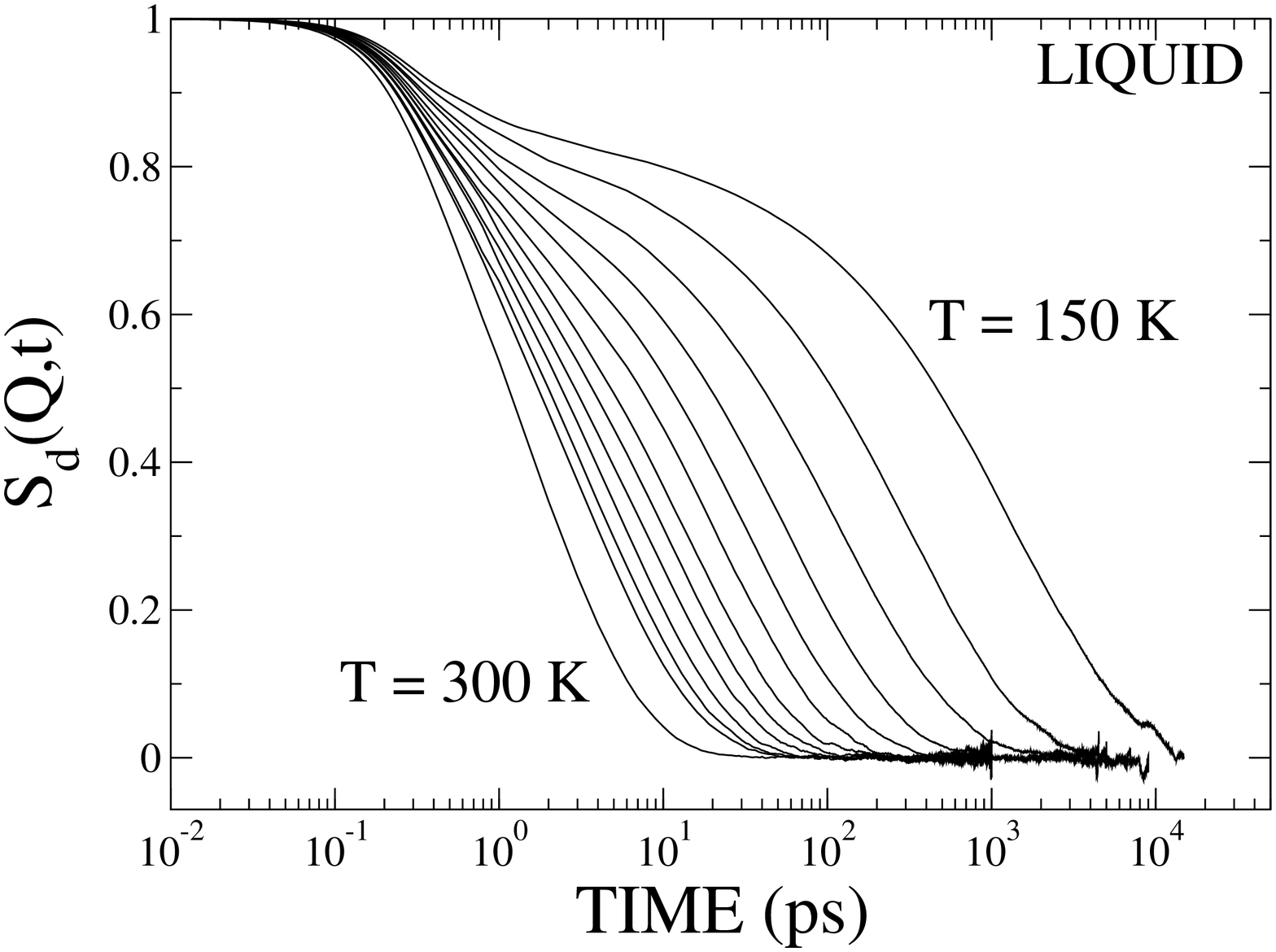}
\caption{\protect\small
Coherent intermediate scattering function $S_{d}(Q,t)$ at $Q = 1.54 \ \mathrm{\AA}^{-1}$
obtained for the plastic phase from $T = $ 130 to 260 K and the liquid phase
from $T = $ 150 to 300 K.
}
\label{figure1}
\end{figure}

MD simulations were performed for 
an orientationally disordered crystal sample of DFTCE
at 14 different temperatures from
$T =$ 130 to 260 K in steps of 10 K
and a supercooled liquid sample at 13 different temperatures from 
$T =$ 150 to the melting temperature $T_{m} \simeq $ 300 K in steps of 10 K.
The same model was used for both phases and it is completely described in~\cite{dftce_model,Affouard_aip03}.

The intermediate scattering function as it can be classically obtained from coherent neutron 
scattering experiments will be mainly considered: 
$S(\vec{Q},t)= \langle
 \rho_{\vec{Q}}(t)\rho_{\vec{Q}}(0) \rangle$
where $\rho_{\vec{Q}}(t)$ is the time-dependent density correlator: 
$\rho_{\vec{Q}}(t) = \sum_{\alpha}b_{\alpha}exp[i\vec{Q}.\vec{r}_{\alpha}(t)]$
where the sum is over all the atoms $\alpha$ of the system.
$b_{\alpha}$ and $\vec{r}_{\alpha}$ are 
the coherent scattering length
and the position of the $\alpha$ atom respectively.
An average over isotropically distributed
$Q$-vectors having the same modulus $Q$
is performed in order to obtain $S(Q,t)$ for a polycrystalline sample. In general, 
$S(Q,t)$ can be expressed as~\cite{Dolling_mp79}:
$S(Q,t)= S_{c}(Q) + S_{d}(Q,t)$
where $S_{c}(Q) = |\langle \rho_{\vec{Q}} \rangle |^{2}$ is the coherent elastic scattering
and 
$S_{d}(Q,t)= \langle \delta \rho_{\vec{Q}}(t) \delta \rho_{\vec{Q}}(0)  \rangle $
where $\delta \rho_{\vec{Q}}(t) =  \rho_{\vec{Q}}(t) - \langle \rho_{\vec{Q}} \rangle$
is the fluctuation of the time-dependent density operator.
For plastic crystals,
$S(Q,t)$ is identical to $S_{d}(Q,t)$ except for $Q$-vectors
corresponding to Bragg peaks
for which the long time limit of $S(Q,t)$ reaches
the non-zero value $S_{c}(Q)$.
This behaviour 
is found at all temperatures and it is associated to the crystalline order of the TDOF and not the 
freezing process of the ODOF.
Therefore, in the following, $S_{d}(Q,t)$ will be preferred to $S(Q,t)$ in order to be able to compare
directly molecular liquids and plastic crystal behaviours.

\begin{figure}[h]
\includegraphics[width=7cm,clip=true]{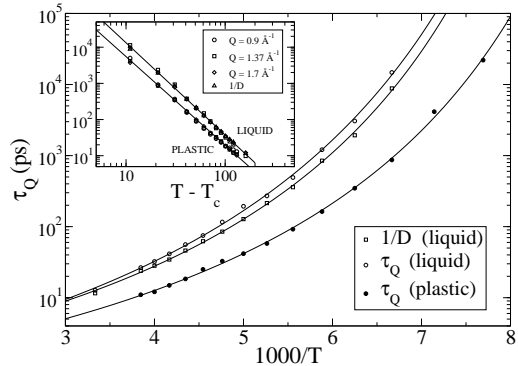}
\caption{\protect\small
Temperature dependence of the relaxation $\tau_{Q}$ at $Q = Q_{max} \simeq 1.27 \ \mathrm{\AA}^{-1}$ for both
the plastic and the liquid phase. Inverse of the diffusion constant $D$ is also indicated.
Solid lines show VFT fits (see text).
Relaxation times $\tau_{Q}$ at $Q = $ 0.9, 1.37 and 1.7 $\mathrm{\AA}^{-1}$ and diffusion constant $D$
($10^{-9}.\mathrm{m^{2}s^{-1}}$) as function of the
rescaled temperature $T - T_{c}$ are displayed in inset.
Solid lines show fits using power law dependence $(T-T_{c})^{\gamma}$.
($T^{l}_{c} \simeq 139 $ K, $\gamma^{l} \simeq 2.55$)
and ($T^{p}_{c} \simeq 129 $ K, $\gamma^{p} \simeq 2.45$) are found
for the supercooled
liquid and the supercooled plastic crystal respectively. Relaxation times are shifted in order
to demonstrate the overlap and the generic power law dependence.
}
\label{figure2}
\end{figure}

\begin{figure}[h]
\includegraphics[width=7cm,clip=true]{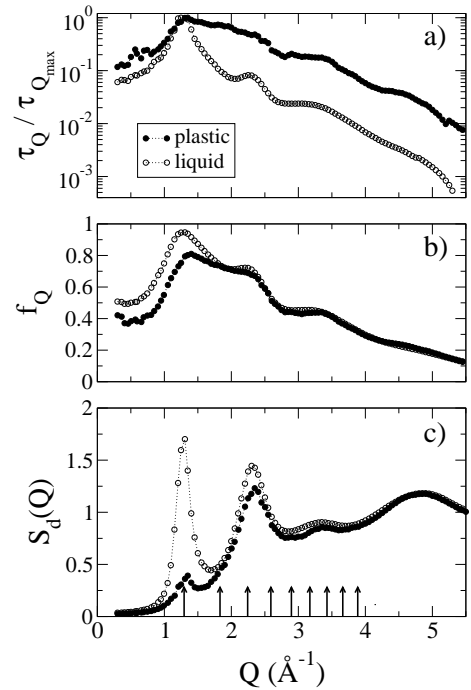}
\caption{\protect\small
$Q$-dependence of different parameters
for the plastic (full circles) and
the supercooled liquid  (circles) phase at the temperature $T = 160 $ K.
a) The $\alpha$-relaxation times $\tau_{Q}/\tau_{Q_{max}}$ 
where $\tau_{Q_{max}}$ is about 345 ps
in the plastic phase and 4122 ps in the liquid phase at this temperature.
b) The non-ergodicity parameter $f_{Q}$ obtained from a fitting procedure
using a von Schweidler law as derived from MCT including second order 
corrections.
c)
The diffuse scattering $S_{d}(Q)$.  First Bragg peaks are also indicated with vertical arrows.
}
\label{figure3}
\end{figure}

Upon cooling, as shown in Fig.~\ref{figure1}, both plastic and liquid $S_{d}(Q,t)$ display
a two-step decay as classically observed 
in supercooled systems~\cite{Kaemmerer_pre97}. 
The long time $\alpha$-relaxation of both plastic and liquid $S_{d}(Q,t)$
is separated from short time regime ($\beta$) by 
a plateau-like region. 
This latter can be associated to the so-called cage effect corresponding to 
the trapping of molecules. 
The nature of this cage is of translational origin
for liquids and of rotational origin (steric hindrance) for plastic crystals~\cite{Affouard_jncs02}.
The height of this plateau is usually called the non-ergodicity parameter and
will be noted $f_{Q}$ in the following.
In order to probe the long time regime,
we also defined the characteristic $\alpha$-relaxation time $\tau_{Q}$ of the rotational dynamics 
as the time it takes for $S_{d}(Q,t)$ to decay from 1
to $0.1$. 
In Fig.~\ref{figure2}, we plot the relaxation time $\tau_{Q}$ 
at $Q = Q_{max}$ corresponding the first diffraction peak (see Fig.~\ref{figure3}c).
In addition, for liquid DFTCE, the translational diffusion constant $D$
has also been calculated from the mean square displacement $\left<r^{2}(t)\right>$.
$\tau_{Q}$ characteristic times and $D$ are found to exhibit
a very similar non-Arrhenius behavior. In order to quantify this similarity,
we used a Vogel-Fulcher-Tammann (VFT) law written as $\tau = \tau_{\infty} \exp [1/K_{VFT}.(T/T_{VFT}-1)]$
where $T_{VFT}$ is the temperature of apparent divergence of $\tau$ and $K_{VFT}$ measures the kinetic fragility.
We particularly obtained $K_{VFT} \simeq $ 0.18, 0.16 and 0.20  and $T_{VFT} \simeq $ 92.5, 90.4 and 86.9 K
from the inverse of the diffusion coefficient, $\tau_{Q}$ of the liquid and  $\tau_{Q}$ of the
plastic crystal, respectively.  It should be noted that results obtained 
from MD calculations are in fair agreement with the parameters  $K_{VFT} = 0.13$ and $T_{VFT} = 70$ K found 
from dieletric experiments of plastic DFTCE~\cite{Kruger_jpcm94}. 

In~\cite{Affouard_prl01}, we suggested that 
some common microscopic mechanisms, relatively well described by MCT, are involved in both 
orientationally disordered crystals and molecular liquids. 
Therefore, in the following, 
we discuss in the MCT framework the wave-vector and temperature dependence of dynamics
for both plastic and liquid DFTCE. 
MCT predictions on the scaling features $\tau_{Q}(T) = A(Q).(T - T_{c})^{-\gamma}$ in the long time decay 
have been carefully checked using a procedure completely described
in~\cite{Affouard_prl01} and results are displayed in Fig.~\ref{figure2}. 
For plastic crystal DFTCE 
it was possible to extract the critical temperature $T^{p}_{c} \simeq 129 $ K and 
scaling exponent $\gamma^{p} \simeq  2.45$
from independent fitting analysis performed in the $\alpha$ regime
at different $Q$. 
Using the same method,
$T^{l}_{c} \simeq 139 $ K and $\gamma^{l} \simeq 2.55$ were found for the supercooled liquid DFTCE.
This good agreement with the $T$-dependence MCT predictions is also confirmed by other tests
in the $\beta$ regime where data were analyzed using a von Schweidler law including the second 
order correction
which will be described elsewhere. $Q$-dependence MCT predictions are presented in the following.
A fundamental property of MCT or MMCT stresses that dynamics are driven by the static density correlator $S(Q)$. 
This has been confirmed in several numerical and experimental studies 
for simple~\cite{Kob_pre1995} or molecular liquids~\cite{Mossa_pre01}. 
In order to check this behaviour in plastic crystals, 
we calculated the static density correlator $S(Q) = S(Q,t=0)$ 
at different temperatures and wave vectors
as well as the non-ergodicity parameter $f_{Q}$, the total prefactors $h^{1}(Q)$ and $h^{2}(Q)$ and the relaxation time $\tau_{Q}$.
A fitting procedure using a von Schweidler law as derived from MCT including second order correction was employed~\cite{Affouard_prl01}.
Similarly to the intermediate scattering function, 
the static correlator can be decomposed
into two components: $S(Q)= S_{c}(Q) + S_{d}(Q)$
where 
$S_{d}(Q)=  \langle |\delta\rho_{\vec{Q}}|^{2} \rangle
= \langle |\rho_{\vec{Q}}| ^{2}  \rangle - |\langle \rho_{\vec{Q}} \rangle |^{2}$
is the diffuse scattering.
Owing the high rotational disorder, plastics crystals yield only a few diffraction peaks
but exhibit intense and highly structured diffusive scattering which provide useful information to 
characterize rotational motions and most probable orientations~\cite{Dolling_mp79}. 
Moreover, remarkable similarities of the structure factor between 
plastic and liquid phases of different systems have been particularly observed~\cite{Fayos_prl96,Dolling_mp79}.
Both $S_{c}(Q)$ and $S_{d}(Q)$ are displayed in Fig.~\ref{figure3}c as function of the wave vector $Q$.
At the largest $Q$-vectors, 
both plastic and liquid $S_{d}(Q)$ merge since the intramolecular structure is probed by the wave vector $Q$.
$\tau_{Q}$ and $f_{Q}$ 
are also shown in Fig.~\ref{figure3}a and b respectively. 
Both plastic and liquid DFTCE $Q$-dependence clearly exhibit very similar features.
In the investigated $Q$-range $[0.3 - 5.5]$ \AA$^{-1}$, $S_{c}(Q)$ shows a few Bragg peaks 
expected for a bcc crystalline structure
with a cell parameter $ a = 6.82 $ \AA$^{-1}$ (experimentally, $ a = 6.98 $ \AA$^{-1}$).
For $S_{d}(Q)$, four broad bumps are localized at 1.27, 2.30, 3.34 and 4.85 \AA$^{-1}$  
both in the supercooled plastic and liquid phases. 
The first sharp diffraction peak of the liquid particularly corresponds to the 
first and most intense Bragg peak of the plastic phase.
Clearly, we see that $f_{Q}$ and $\tau_{Q}$ obtained for plastic or liquid phases are mainly correlated with 
$S_{d}(Q)$ as predicted by MCT. The total prefactors not displayed in Fig.~\ref{figure3}
exhibit the same behaviour. 
The non-ergodicity parameter or the total pre-factor obtained from $S(Q)$ of the plastic phase also shows
a modulation with Bragg peaks of the elastic scattering $S_{c}(Q)$ (not shown in the present Letter).
It should be mentioned that colliding infinitely thin needles distributed on a lattice are often considered
as prototype of glassy crystals~\cite{Renner,Schilling_epl03}.
Two-step relaxation processes are observed and an ideal glass transition can be obtained
when the length of the rods (the analog of the temperature for these systems)
reaches a critical value.
If the thickness of the rods are chosen infinitely thin,
all static correlations vanish and thus,
following MCT predictions, no glass transition is expected~\cite{Renner,Schilling_epl03}.
The present work clearly illustrates the strong difference between real plastic crystals
and the colliding needles on the origin of their respective ideal glass transition.

In this Letter, collective dynamics of both supercooled plastic crystal and supercooled liquids
states of DFTCE have been investigated using MD simulations. 
From the calculation of the total intermediate scattering
functions $S(Q,t)$ at different wave vectors $Q$ and temperatures $T$,
our results confirm the strong analogy between two systems made of the same molecules but 
for which dynamics are controlled by different degrees of freedom.
The authors hope that the present investigation could contribute to develop a theoretical framework
for orientationally disordered crystal based on MCT

The authors wish to acknowledge the use of the facilities of the IDRIS (Orsay, France)
and the CRI (Villeneuve d'Ascq, France)
where calculations were carried out.
This work was supported by the INTERREG III (FEDER)
program
(Nord Pas de Calais/Kent).


\begin{thebibliography}{24}
\bibitem[{dft()}]{dftce_model}
\bibinfo{journal}{\protect MD calculations were performed using the DL\_POLY
  package (W. Smith and T. Forester, CCLRC, Daresbury Laboratory) on a system
  of $N = 686$ ($7 \times 7 \times 7$ bcc crystalline cells) DFTCE molecules.
  Each molecule is described by its 8 atoms and considered as rigid. The system
  is made of a constant 50\% ratio of both trans and gauche conformers. The
  very slow trans-gauche transformation has been neglected in our simulation.
  Molecules only interact through a Buckingham short range atom-atom potential
  whose parameters are taken from litterature and given
  in~\cite{Affouard_aip03}. The electrostatic contributions have been neglected
  since the DFTCE molecule possesses a weak dipolar moment. Both structure and
  dynamics of the DFTCE plastic phase obtained from MD were found in good
  agreement with X-rays diffraction, Neutron Spin Echo (NSE) and NMR
  experimental results. Newton's equations of motion were solved with a time
  step of $\Delta t = 5$ fs. We worked in the NPT statistical ensemble with
  periodic boundaries conditions where the simulating box is allowed to change
  in size and shape.}
\end{thebibliography}
\end{document}